\documentclass[aip,jcp,reprint,amsmath,amsfonts,amssymb,a4paper,eqsecnum,superscriptaddress]{revtex4-1}

\usepackage{graphicx}
\usepackage{float}
\usepackage{color}

\newcommand{\void}[1]{}
\renewcommand{\nu}{2\epsilon-\Delta} 

\begin{document}

\title{Thermal activation at moderate-to-high and high damping: finite barrier effects and
  force spectroscopy.}

\author{J.~J. Mazo}
\affiliation{Departamento de F\'{\i}sica de la Materia Condensada and Instituto de Ciencia de Materiales de Arag\'on, CSIC-Universidad de Zaragoza, 50009 Zaragoza, Spain}
\author{O. Y. Fajardo}
\affiliation{Departamento de F\'{\i}sica de la Materia Condensada and Instituto de Ciencia de Materiales de Arag\'on, CSIC-Universidad de Zaragoza, 50009 Zaragoza, Spain}
\author{D. Zueco}
\affiliation{Departamento de F\'{\i}sica de la Materia Condensada and Instituto de Ciencia de Materiales de Arag\'on, CSIC-Universidad de Zaragoza, 50009 Zaragoza, Spain}
\affiliation{Fundaci\'on ARAID, Paseo Mar\'{i}a Agust\'{\i}n 36, E-50004 Zaragoza, Spain}

\date{\today}

\begin{abstract}
We study the thermal escape problem in the moderate-to-high and high
damping regime of a system with a parabolic barrier. We present a
formula that matches our numerical results accounting for finite
barrier effects, and compare it with previous works. We also show results
for the full damping range. We quantitatively study some aspects on
the relation between mean first passage time and the definition of a
escape rate. To finish we apply our results and considerations in the
framework of force spectroscopy problems. We study the differences on
the predictions using the different theories and discuss the role of
$\gamma \dot{F}$ as the relevant parameter at high damping.
\end{abstract}

\maketitle

\section{Introduction}

Seventy years ago, Kramers~\cite{Kramers1940} proposed an equation for
the thermal escape of a Brownian particle out of a metastable
well. This problem is considered an issue of fundamental interest and
it has received great attention from an impressive number of
scientific areas~\cite{Hanggi1990a, Melnikov1991a,Pollak2005a}.  We
present here a minor contribution to this vast field: a detailed
numerical study, by using Langevin dynamics simulations, of the
activation problem in the moderate-to-high and high damping regime of
the system. 

This work is motivated by a recent numerical study of the Kramers
problem at low damping~\cite{Mazo2010}.  There, important finite
barrier effects were described, a very slow convergence to the
infinite barrier limit of the system was reported and the accuracy of
the Drozdov-Hayashi (DH) theory~\cite{DH1999} was proved. As it is shown
there, an accurate theory is of fundamental interest to understand
experiments in this physical regime.

On the other side of the same coin, the high damping limit of the problem is
found. This is the typical case for many of the current
friction~\cite{Gnecco00,Sang01,Evstigneev04,Evstigneev06,Barel10} and
biological-physics works~\cite{Hyeon2003,Dudko2003,Hummer03},
where escape rate is the fundamental concept to understand force
spectroscopy experiments. In many of these works the Kramers result for the
escape rate at high damping:
\begin{equation}
\label{Kramers_hd}
r_{_{KHD}} = \frac{\omega_a \omega_b}{2 \pi \gamma} \, {\rm e}^{- \Delta U/k_B T},
\end{equation}
appears as the theoretical starting point to understand experimental results.

\begin{figure}[]
    \centering{\includegraphics[width=0.25\textwidth]{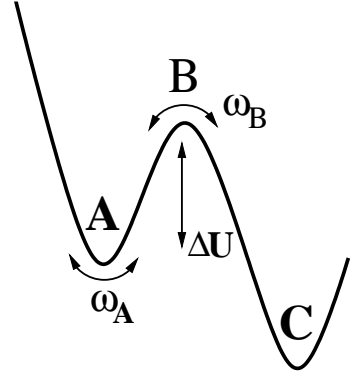}}
    \caption{Potential profile. Due to thermal fluctuations, a
      particle sited in state $A$ is able to overcome a barrier
      $\Delta U$ and escape to state $C$.}
  \label{fig:pot}
\end{figure}

In his famous work\cite{Kramers1940}, Kramers also derived an expression for the escape
rate in the moderate to high damping limit,
\begin{equation}
\label{Kramers}
r_{_{KMHD}} = k_{_{KMHD}} \times \frac{\omega_a}{2 \pi} \, {\rm e}^{- \Delta U/k_B T},
\end{equation}
where the damping dependent prefactor is given by 
$k_{_{KMHD}} (\gamma/\omega_b) =
\left[1+(\gamma/2\omega_b)^2\right]^{1/2}-(\gamma/2\omega_b)$.
There, $\omega_a$ and $\omega_b$ are related with the potential
curvature at the well and the barrier respectively, see
Fig.~\ref{fig:pot}, $\Delta U$ stands for the potential barrier and
$\gamma$ for the damping of the system. For high damping,
$\gamma/\omega_b \gg 1$, we have $k_{_{KHD}}:=\omega_b/\gamma$ for
the prefactor and we recover the extensively used result of
Eq.~(\ref{Kramers_hd}).  The given results were obtained for a parabolic
barrier~\cite{comment1} in the so-called infinite barrier regime of
the system, where $\Delta U/k_B T \gg  1$.

Another important concept in thermal activation theory is the mean
first passage time (MFPT). This is defined as the mean time
$\langle t \rangle$ for particles injected with zero velocity at point
$a$, to reach a point $b$, where they are absorbed. In the overdamped
limit, given the potential $U(x)$, the MFPT can be exactly computed
as~\cite{Hanggi1990a}
\begin{equation}
\langle t \rangle_{_{HD}}=\frac{m \gamma}{k_B T} \int_a^b dy \, {\rm
  e}^{U(y)/k_BT} \int_{-\infty}^y dz \, {\rm e}^{-U(z)/k_BT}.
\label{mfpt}
\end{equation}

There exist an important literature studying the relation between
transition rates and mean first passage
times~\cite{Hanggi1990a,Pollak2005a,Muller97,Reimann99}. It has been
usually assumed, and theoretically proved, that there is a close
relation between both quantities and in fact
\begin{equation}
r=\frac{1}{\langle t \rangle},
\label{r_t}
\end{equation}
whenever the injection and adsorbing points correspond to two
potential minima separated by a barrier. Thus, Eqs.~(\ref{mfpt}) and
(\ref{r_t}) allow for computing the exact escape rate in the
overdamped limit of the system and for comparing to the Kramers
infinite barrier and other escape rate results in this physical
regime.

We devote section II of the paper to this end where a critical review
of existing theories is presented. There, we propose a relatively
simple equation, Eq.~(\ref{rmhd}), to compute the escape rate for
any barrier and damping in the moderate-to-high and high damping
domains. Then, in section III we study the time to reach the barrier
problem. This time, $\langle t \rangle_M$, differs from the previously
calculated $\langle t \rangle$ by a factor that changes from 1 to 1/2
when damping increases, and it is the relevant quantity in some
situations.

Armed with our previous results, we attack in section IV the so-called
{\em force spectroscopy} problems. There, the system is continuously
biased by an external field and the escape field probability
distribution function is recorded. We compare results for the mean
escape field based in different theories. We stress the differences
between results for field to reach the next metastable state and field
to reach the barrier. We also show that damping times field ramp is
the relevant parameter in the overdamped limit, and that there exists
a value of this parameter beyond which important nonequilibrium
effects dominate the problem. We finish by showing results in the full
damping spectrum.

\section{Model and results}

We use Langevin dynamics simulations to compute the mean time for a
particle with a given initial condition to reach a defined point
beyond the barrier. To be precise, we will be first interested in
computing the mean time for a particle sited with zero velocity in the
minimum of a metastable well $A$ to reach for first time the next
available potential minimum $C$, see Fig.~\ref{fig:pot}. For us this
will be the definition of the relevant mean first passage time (MFPT)
$\langle t \rangle$ of the problem.

To be definite we study the dynamics of a Brownian particle in a
metastable potential:
\begin{equation}
m\ddot{x}+  m \gamma \dot{x}=-\frac{dV}{d x}+\xi(t),
\label{langevin}
\end{equation}
where for the potential we use
\begin{equation}
V(x)=V_0 (1-\cos{x})-F x,
\label{pot}
\end{equation}
and $\xi(t)$
is the stochastic force describing the thermal fluctuations. Here we
consider white thermal noise, $\langle \xi(t) \rangle = 0$ and
$\langle\xi(t)\xi(t')\rangle= 2 m \gamma k_B T \delta(t-t')$.

The tilted sinusoidal potential describes a particle in a periodic
potential subjected to an external field, a situation of interest in
many fields. In particular, Eq.~(\ref{langevin}) is experimentally
realized by a biased Josephson junction, and the high damping regime
can be addressed with resistively shunted tunnel junctions or with
superconducting-normal-superconducting ones\cite{Barone}.

We compute the escape rate of a particle out of the metastable well
(see Fig.~\ref{fig:pot}) as a function of the damping of the system and
for different values of the $\delta := \Delta U/k_BT$ ratio, which is
the most relevant parameter of the problem. Some authors, using
different approaches, have addressed the problem of including finite
barrier effects in the
theory~\cite{Pollak93,Melnikov93,Drozdov99}. This is an interesting
issue since in many cases escape occurs at relatively small barriers.
In our model system, the energy barrier depends on the normalized
external force $f=F/V_0$ as $\Delta U=2V_0
\left[\left(1-f^2\right)^{1/2}-f\cos^{-1} f \right]$.
Another important parameter in the theory is the value of the
curvature in the metastable minimum ($\omega_a$) and the barrier
($\omega_b$). In this case $\omega_a=\omega_b=\left(V_0/m\right)^{1/2} \left(1-f^2\right)^{1/4}$.

\begin{figure}[]
    \centering{\includegraphics[width=0.47\textwidth]{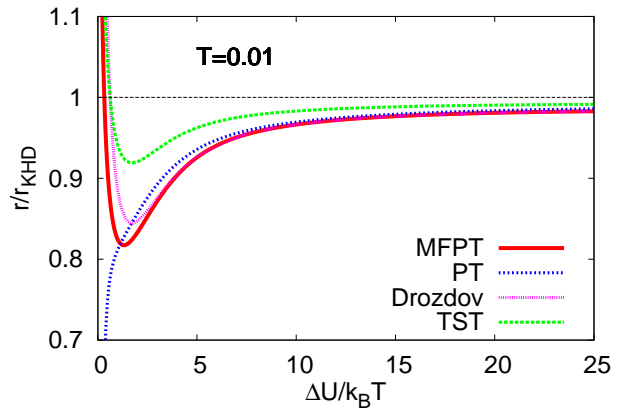}}
    \caption{(color online) Theoretical estimations of the escape rate
      in the high damping limit normalized by the Kramers result. Red
      line shows the exact mean first passage time result which
      accounts for the finite barrier effects in the system. We also
      show PT, Drozdov and exact TST results (see text and appendix).}
  \label{fig:hdt}
\end{figure}

We have performed molecular dynamics simulations of Langevin
equation~(\ref{langevin}).  The
equations were numerically integrated using the stochastic Runge-Kutta
algorithm~\cite{RK} and the MFPT was computed typically over 10000
realizations. As in our previous work~\cite{Mazo2010}, we have used
$m=0.35$ and $V_0=0.155$. In most of the simulations we use
$k_BT=0.01$ and change $F$ and $\gamma$.\cite{comment2}

\subsection{High damping}
First, we compare in Fig.~\ref{fig:hdt} exact results from the MFPT
theory [Eqs.~(\ref{r_t}) and~(\ref{mfpt})] to the Kramers infinite
barrier one, Eq.~(\ref{Kramers_hd}). We show also the Pollak and
Talkner~\cite{Pollak93} (PT) and Drozdov~\cite{Drozdov99} results,
which include finite barrier corrections to the escape rates. As a
reference we also plot the $r_{_{TST}}^{exact}$ result~\cite{comment3}
[Eq.~(\ref{tst_e}) in the appendix]. These results were proposed for
the high damping limit of the system. The figure gives a quantitative
estimation of the accuracy of the Kramers high damping result for
every barrier.  As it can be seen, finite barrier effects are not very
severe (below $10\%$) for normalized barriers above 5. However some
care have to be taken when considering effects involving barriers
below this value. Figure also shows the convergence of the escape rate
to the Kramers infinite barrier result.

As we also see in Fig.~\ref{fig:hdt}, PT theory gives a {\em simple}
expression very close to the exact result except for small
barriers. Drozdov gives an even better approach, however its final
equation is {\em more complex} and only valid for very high values of
the damping. PT approach fails at low barriers and Drozdov is a good
approximation only in the overdamped limit.

\subsection{Moderate-to-high damping}
We propose below an expression for the escape rate which gives a good
approximation for any barrier height and
damping. Equation~(\ref{mfpt}) is exact in the overdamped limit of the
system. However there is no result for the MFPT at other values of the
damping of the system.

\begin{figure}[]
    \centering{\includegraphics[width=0.47\textwidth]{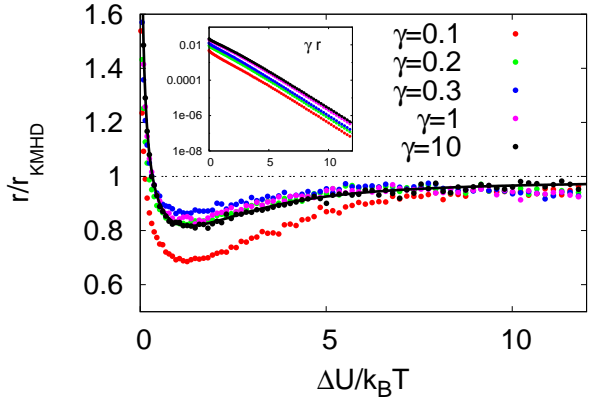}}
    \caption{(color online) Numerical estimations of the escape rate
      in the moderate-to-high damping regime normalized by the Kramers
      result of Eq.~(\ref{Kramers}). Black line shows the exact mean first
      passage time result for the high damping case.}
  \label{fig:hd}
\end{figure}

Our proposal is based in observing our numerical results for the
escape rate at different values of damping as a function of the
normalized barrier $\Delta U/k_B T$. Figure~\ref{fig:hd} shows results
for $\gamma r$ (inset) and $r/r_{_{KMHD}}$ [note that, different from
  Fig.~\ref{fig:hdt}, now we normalize the rate with respect to
  Eq.~(\ref{Kramers})] as a function of $\Delta U/k_B T$ for $T=0.01$
and $\gamma=0.1, 0.2, 0.3, 1$ and 10. There, we see first that our
numerical results for high damping agree the theoretical prediction
from the MFPT expression, Eq.~(\ref{mfpt}).

Figure~\ref{fig:hd} also shows that, except for the $\gamma=0.1$ curve
(we will show below that this value of damping corresponds to the
moderate damping region where $r(\gamma)$ gets a maximum), all data
approximately lie in a same curve when compared to the $r_{_{KMHD}}$
prediction. Thus, we propose the following simple extension of the
MFPT results of Eq.~(\ref{mfpt}) to the moderate-to-high damping
region:
\begin{equation}
r_{_{MHD}}=\frac{\gamma}{\omega_b} \times k_{_{KMHD}}\left( \frac{\gamma}{\omega_b} \right) \times
\frac{1}{\langle t \rangle_{_{HD}}}.
\label{rmhd}
\end{equation}
The formula recovers the correct $r_{_{HD}}=1/\langle t
\rangle_{_{HD}}$ result for $\gamma/\omega_b \gg 1$ and incorporates the
$\gamma$ dependence suggested in Fig.~\ref{fig:hd}.  As we will see,
this expression gives account of our numerical results in the
moderate-to-high damping regime of the problem in a simpler and better
way than any previous approach.

\begin{figure}
    \centering{\includegraphics[width=0.47\textwidth]{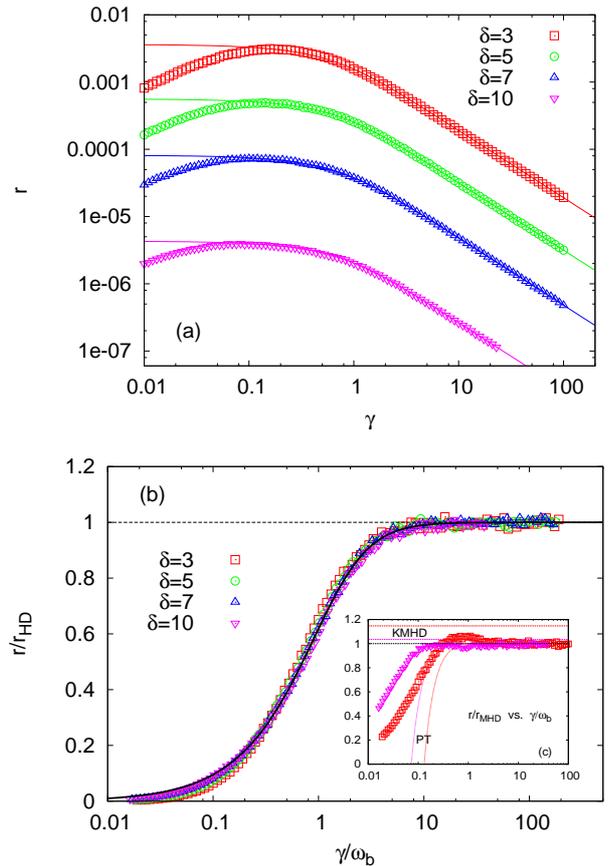}}
    \caption{(color online) Escape rate at different $\gamma$ for
      $\delta:=\Delta U/k_BT=$3, 5, 7 and 10, and prediction of
      Eq.~(\ref{rmhd}). (a) Escape rate as a function of $\gamma$. (b)
      Escape rate normalized by the high damping result as a function
      of $\gamma/\omega_b$. (c) Escape rate normalized by the result
      of Eq.~(\ref{rmhd}) for $\Delta U/k_BT=3$ and 10 and prediction of the
      Kramers (dashed lines) and PT (dotted lines) theories.}
  \label{fig:hd_g}
\end{figure}

In order to evaluate the accuracy of Eq.~(\ref{rmhd}), we have
performed numerical simulations of the escape rate of the system as a
function of $\gamma$ for different values of the barrier. In
particular, Fig.~\ref{fig:hd_g} shows our numerical results obtained
for four different values of the $\Delta U/k_BT$ ratio (3, 5, 7 and
10). For better understanding, we plot results in three different
ways. Figure.~\ref{fig:hd_g}(a) shows the rate versus the damping. As
we see, the theoretical expression is valid for values of $\gamma$
larger than 0.1 approximately. This value is at the lower limit of the
moderate-to-high damping region. Smaller values of $\gamma$ sets in
the moderate-to-low damping regime that we studied
previously~\cite{Mazo2010}. These kinds of plots have been extensively
used in the literature, however due to the log scale dependence of the
rate it is not convenient to use them to evaluate the precision of the
theoretical equations.

\begin{figure}[]
    \centering{\includegraphics[width=0.47\textwidth]{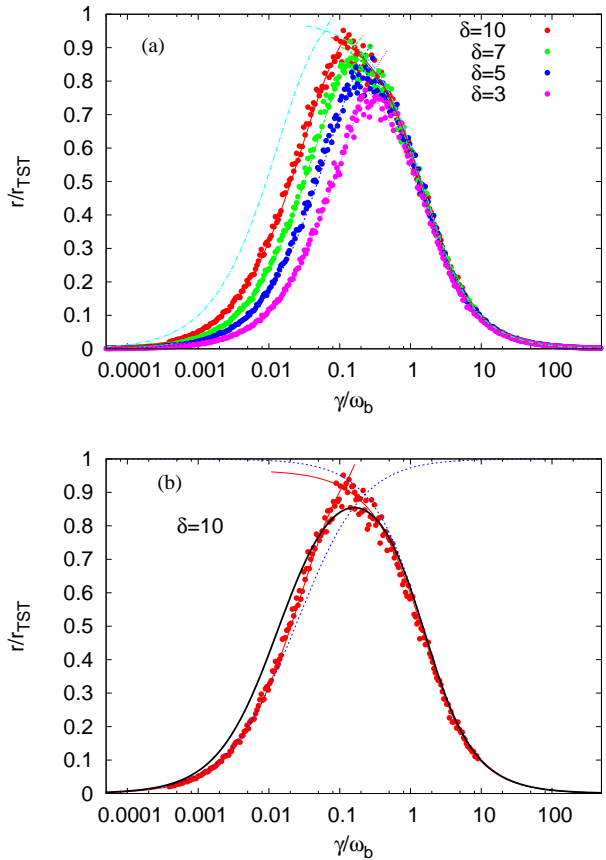}}
    \caption{(color online) (a)Theoretical (lines) and numerical
      (points) escape rates as a function of the normalized damping
      parameter for different values of the barrier over temperature
      ratio. Higher lines are theoretical predictions for $\Delta
      U/k_BT=20$. (b) Different theoretical approaches to the escape
      rate and comparison to our numerical results, for $\Delta
      U/k_BT=10$. Red lines stands for Drozdov-Hayashi (low damping)
      and for our $r_{_{MHD}}$ (high damping) results. Blue dotted
      lines stands for the BHL approach (low damping) and the KMHD
      one. Black line shows the simplest interpolation equation for
      the full damping range (see text).}
  \label{fig:rgw}
\end{figure}

Figure~\ref{fig:hd_g}(b) plots $r/r_{_{HD}}$ as a function of
$\gamma/\omega_b$ to check the validity of our ansatz in
Eq.~(\ref{rmhd}). Plotted in this way, all the curves approximately
collapse in a single function given by
$(\gamma/\omega_b)\times k_{_{KMHD}}(\gamma/\omega_b)$. As expected,
deviations appears for the lowest values of damping.

To finish, we present in Fig.~\ref{fig:hd_g}(c) the direct comparison
of our numerical results to Eq.~(\ref{rmhd}) and PT theory for
$\delta=3$ and $\delta=10$. In this figure we plot the escape rate
normalized by our theoretical expression, $r/r_{_{MHD}}$ versus
$\gamma/\omega_b$.  We conclude that Eq.~(\ref{rmhd}) reproduces our
numerical results in a better way than previous theories for any
barrier and in the full moderate-to-high damping regime.

As we have seen, our numerical results are well fitted by
Eq.~(\ref{rmhd}) for damping values above $\gamma \sim 0.2$ in our
units. At this intermediate values of the damping parameter we are
into the so-called turnover region, which characterizes the transition
between the moderate-to-low and the moderate-to-high damping
regimes. There, the classical transition state theory establishes
$r_{_{TST}}=(\omega_a/2\pi)\exp{(-\Delta U / k_B T)}$.

\subsection{Full damping}
In order to supplement the information given in previous works,
Fig.~\ref{fig:rgw}(a) shows the escape rate divided by the
$r_{_{TST}}$ result for a wide range of damping values with special
emphasis in the turnover region. As expected, when plotted as a
function of the normalized damping $\gamma/\omega_b$ all curves
collapse one each other only in the high damping limit. As barrier
increases, the maximum moves to lower values of $\gamma/\omega_b$. In
any case rate approaches very slowly to the infinite barrier
limit. The figure also shows the theoretical results studied in
Ref.~\onlinecite{Mazo2010} and here. We see that the combination of
the Drozdov-Hayashi~\cite{DH1999} result for moderate-to-low damping
values and our extension of the mean first passage time result for
moderate-to-high ones, give a fairly good approximation to the
numerically computed result for all damping of the system.

Fig.~\ref{fig:rgw}(b) shows the results for $\Delta U/k_B T=10$ and
also plots three other theoretical approaches. On the one hand we use the
Kramers moderate to high damping result and the
Buttiker-Harris-Landauer~\cite{bhl} (BHL) one. We see that the latter
deviates importantly in the moderate damping region of the system. To
finish we also plot the result of the simplest analytical expression
which approach the escape rate for all damping, which is given by the
following interpolation formula\cite{Pollak89,Hanggi1990a,Linkwitz92}:
\begin{equation}
r_{\rm simple}^{-1}=r_{_{KLD}}^{-1}+r_{_{KMHD}}^{-1},
\label{simple}
\end{equation}
where for a linear-cubic barrier we have $r_{_{KLD}}=(7.2\gamma/2\pi) (\Delta
U/k_BT){\rm e}^{-\Delta U/k_BT}$ and $r_{_{KMHD}}$ is given by
Eq.~(\ref{Kramers}). This is a simple expression which, as seen in the
figure, gives a first glance good idea of the escape rate behavior in
the full damping spectrum of the system.

\begin{figure}[]
    \centering{\includegraphics[width=0.47\textwidth]{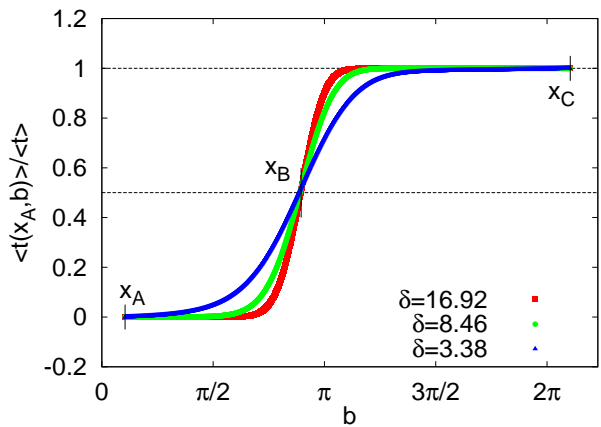}}
    \caption{(color online) Figure shows $\langle t(a,b) \rangle$ as a
      function of $b$ for $a=x_{A}$ and $b \in (a,a+2\pi)$. For
      comparison, results are normalized with respect to $\langle t
      \rangle=\langle t(x_A,x_C) \rangle$ for each case. We set
      $F=0.05$ and plot results for three different temperatures,
      resulting in different $\Delta U/k_BT$
      ratios. $x_{A}=\sin^{-1}(F/V_0)$, barrier $x_{B}=\pi-x_{A}$ and
      $x_{C}=x_{A}+2\pi$ are marked by vertical segments.}
  \label{fig:MFPT_x}
\end{figure}

\section{Time to reach the barrier}

The general expression for the MFPT, Eq.~(\ref{mfpt}), depends on the
injection, $a$, and absorption, $b$, points. It is interesting to
remark that the adequate choice for both numbers in order to keep
Eq.~(\ref{r_t}) valid usually relies on relatively vague criteria
since for particles starting in a metastable well {\em far enough} the
barrier and adsorbed in another well {\em well beyond} the barrier,
the dependence on $a$ and $b$ is negligible. Thus $a$ and $b$ should
be both far from the barrier. Equation~(\ref{mfpt}) allows for a
direct study of the $\langle t(a,b) \rangle$
dependence. Figure~\ref{fig:MFPT_x} explores $\langle t(a,b) \rangle$
as a function of $b$ for $a=x_{A}$ (the potential minimum position)
and different values of the model parameters. As it can be seen, the
MFPT changes importantly in the barrier region. As expected the time
to reach the next potential minimum approximately doubles the time to
get the maximum. A similar study can be done to know the dependence of
the MFPT on the injection position of the system, $a$.

\begin{figure}[]
    \centering{\includegraphics[width=0.47\textwidth]{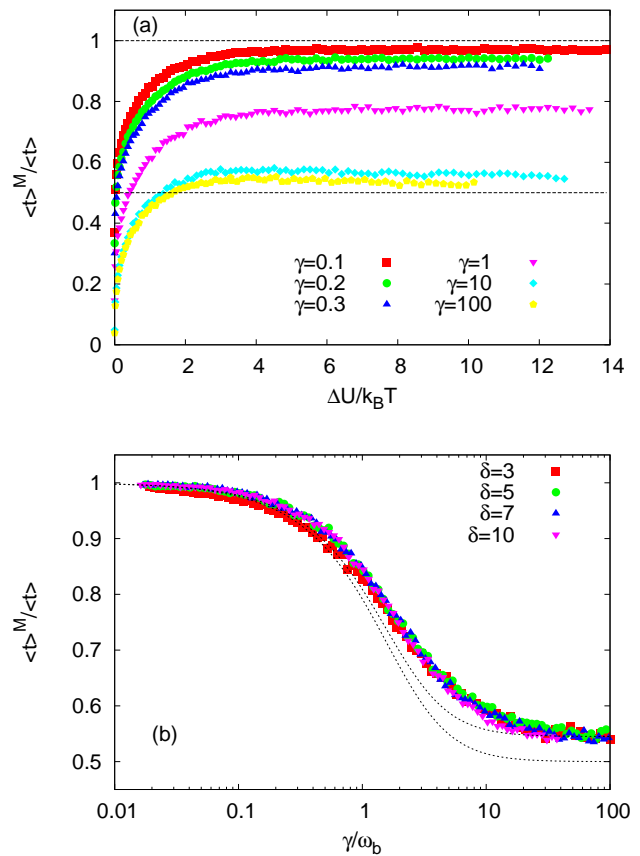}}
    \caption{(color online) $\langle t \rangle^M/\langle t \rangle$
      for different values of $\Delta U/k_BT$ and $\gamma$. In (b) we
      also show (lines) the function proposed in Eq.~(\ref{rmhd_M})
      for $\epsilon_{_{HD}}=0.5$ and 0.545.}
  \label{fig:comp}
\end{figure}

There exist problems where the relevant physical magnitude is the mean
time spent by the system to reach the maximum of the potential barrier
(the transition state $B$ in Fig.~\ref{fig:pot}). We call here this
time $\langle t \rangle^{M}$ to distinguish it from $\langle t \rangle$,
defined above as the time to reach state C beyond the barrier.

Let us introduce the factor $\epsilon:=\langle t \rangle^{M} /\langle
t \rangle$ for the ratio between both times. As can be seen in
Fig.~\ref{fig:MFPT_x}, a rough estimation in the high damping limit is
that $\epsilon_{_{HD}} \simeq 0.5$, showing that a particle in the
maximum has a probability close to one half to cross the barrier and
close to one half to move backwards to the original well. However, this
is not exactly the case and as it is well known~\cite{Muller97} the so
called stochastic separatrix is sited beyond the barrier and differs
from the deterministic separatrix. There, interesting dynamic effects
appear in the Brownian dynamics of the
system~\cite{Fiasconaro05,Fiasconaro10}.

The value of $\langle t \rangle^{M}$ can be computed in the overdamped
limit from Eq.~(\ref{mfpt}) fixing absorption limit $b$ at the
potential barrier position. We have numerically explored how
$\epsilon$ depends on the normalized barrier height $\Delta U/k_BT$
and damping of the system. We should expect a weak dependence on the
barrier at high enough values of $\Delta U/k_BT$. With respect to the
damping, $\epsilon$ changes smoothly from a value close to 0.5 for
high damping to 1 for small damping.  At small damping any particle
that reaches the maximum has enough energy to overcome the barrier and
rapidly slides down to the next potential minimum.

Figure~\ref{fig:comp}(a) plots $\epsilon$ as a function of $\Delta
U/k_BT$ at different values of the damping $\gamma$.
Figure~\ref{fig:comp}(b) shows $\epsilon$ as a function of the
normalized damping $\gamma/\omega_b$ at different values of the
normalized barrier $\Delta U/k_BT$. We can see that $\epsilon$ depends
very weakly on the barrier except for values of $\Delta U \leq 3
k_BT$. In which respects to the damping dependence, $\epsilon$ changes from
$\epsilon_{_{HD}} \simeq 0.55$ at our parameters for $\gamma > 10$ to
$\epsilon \simeq 1$ for $\gamma<0.1$.

In a similar way that for evaluating escape rates, it would be useful
to get an analytical expression to computing the mean time to reach the
maximum at different values of the parameters of the system. From
Fig.~\ref{fig:comp}(b) we see that for every barrier we can write
$\langle t \rangle^{M}=f(\gamma/\omega_b)\langle t \rangle$. Following
the ansatz used to derive Eq.~(\ref{rmhd}) we now propose~\cite{comment4}
\begin{equation}
\frac{1}{\langle t \rangle^{M}_{_{MHD}}}=
\frac{\gamma}{\omega_b} \epsilon_{_{HD}} \times
k_{_{MHD}}\left( \frac{\gamma}{\omega_b} \epsilon_{_{HD}} \right) \times
\frac{1}{\langle t \rangle_{_{HD}}^{M}}
\label{rmhd_M}
\end{equation}
for the case of a parabolic barrier. There $\epsilon_{_{HD}}=\langle t
\rangle^{M}_{_{HD}} /\langle t \rangle_{_{HD}}$ is computed from
Eq.~(\ref{mfpt}). Figure~\ref{fig:comp} compares this equation to the
numerical results for two possible values of $\epsilon_{_{HD}}$, 0.5
and 0.545. Though agreement is worse than that for the escape rate case, to
our knowledge this is the first attempt to propose an expression for the
$\langle t \rangle^{M}(\gamma)$ dependence.

\section{Force spectroscopy}

\begin{figure}[]
    \centering{\includegraphics[width=0.47\textwidth]{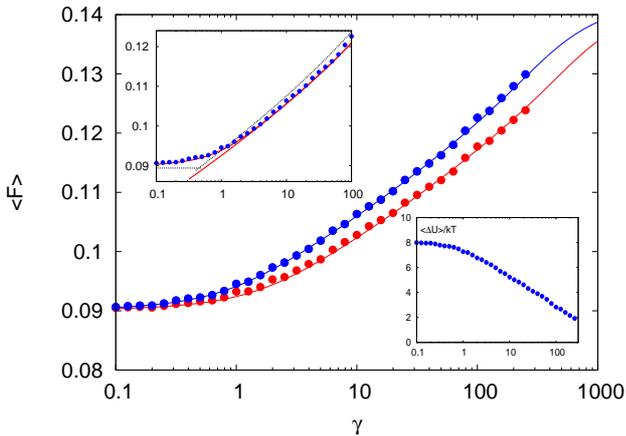}}
    \caption{(color online) Mean force $\langle F \rangle$ as a
      function of the damping $\gamma$ for first passing the barrier
      top (red) and the bottom of the potential well (blue) in a force
      spectroscopy simulation at $T=0.01$ and with force ramp
      2.5$\times$10$^{-7}$. Lines are for theoretical predictions
      based in Eqs.~(\ref{rmhd}) and~(\ref{rmhd_M}).  Top inset:
      comparison to different simple theories. Bottom inset:
      normalized barrier at the different mean escape force
      values. For our parameters, the barrier disappears at
      $F=V_0=0.155$.}
\label{fig:mM}
\end{figure}

In a typical force spectroscopy experiment the probability
distribution function of the escape force $P(F)$ is measured
performing many experiments where force is continuously increased at a
given rate $\dot{F}$. From these results the mean escape force
$\langle F \rangle$ and its standard deviation can be trivially
computed. Such $P(F)$ can be easily related to the escape rates $r(F)$
as~\cite{Fulton1974}
\begin{equation}
P(F)=r(F) \left( \frac{dF}{dt} \right)^{-1} \left( 1- \int_0^F P(u) du \right).
\label{eq:P(F)}
\end{equation}
Alike, escape rates can be computed from measured $P(F)$. In this
section we will show results of the mean escape force in our system for
different parameter conditions and compare to available theories.

As in the case of the escape rate and its relation with the MFPT, it
is possible to introduce different definitions of escape force. In our
simulations we start at zero force with particles in a metastable
state (state A in Fig.~\ref{fig:pot}) and increase the force at a
given ramp. Then we define the escape force as the mean force for which
particles first reach the next metastable state (state C). However,
in some circumstances the relevant escape force can be related with a
different state, for instance the potential barrier (state B). In the
first case we say that a particle has escaped if it has reached a new
potential minimum, in the second if it has just passed the potential
maximum.

Figure~\ref{fig:mM} shows the results of the mean escape force
computed for both cases. As expected, $\langle F \rangle$ increases
with the damping and both curves overlap for moderate damping and
differ at larger ones. Figure also shows the theoretical results
obtained from Eq.~(\ref{eq:P(F)}) with $r(F)$ computed as
Eqs.~(\ref{rmhd}) and (\ref{rmhd_M}), in this last case using
$\epsilon_{HD}=0.545$. As expected, such curves give account of the
numerical results in all the studied range.  We want to remark that
the difference between both curves at high damping is due to the
factor close to 0.5 between the MFPT for reaching states B or C. Thus,
a factor of 0.5 in the prefactor of the rates has a measurable effect
in the measured critical force value, see Fig.~\ref{fig:mM}.

We also compare our results with predictions from Eqs.~(\ref{Kramers})
and (\ref{Kramers_hd}) and analytical results by Garg~\cite{Garg}, see
appendix. Based on the small error found in the estimation of the
system escape rates, we can advance that even the simplest theories
give fairly good results. Such analysis is made in the top inset of
Fig.~\ref{fig:mM} where the theoretical results obtained from the
Kramers high damping and the Kramers moderate-to-high damping results
(red lines) and the first order Garg theory (black dashed lines) for
high and moderate damping are shown.

It is interesting to remark that in this numerical experiment escape
happens at moderate to low values of the barrier over temperature
ratio (see bottom inset in Fig.~\ref{fig:mM}). Thus we see that
results obtained from the infinite barrier limit give reasonable
estimations for physical processes involving such small barrier
values.

Next, we are going to discuss an important issue, regarding the
validity of the available theories at very high damping values or
force rates, where strong nonequilibrium effects plays a role. At high
damping escape events are very rare and, if the force is varied fast
enough the critical force of the model (for which barrier disappears)
is reached before particles have reached the barrier. Then we can say
that the escape problem is somehow ill-defined.

\begin{figure}[]
    \centering{\includegraphics[width=0.47\textwidth]{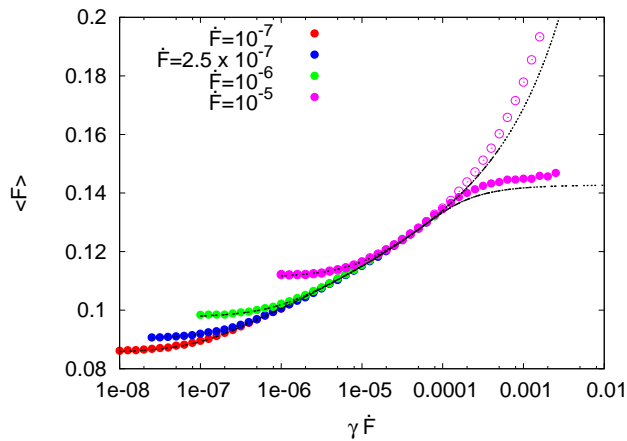}}
    \caption{(color online) Mean escape force as a function of $\gamma
      \dot{F}$ for different force ramps. Open and solid symbols state
      for two different escape force definition criteria (see text).
      Lines are theoretical predictions for both cases.}
  \label{fig:fc_gr}
\end{figure}

\begin{figure}[]
    \centering{\includegraphics[width=0.47\textwidth]{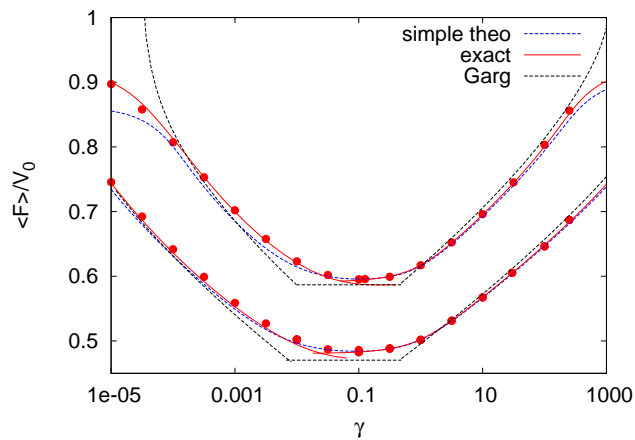}}
    \caption{(color online) Mean escape force as a function of
      $\gamma$ for two different force ramps $\dot{F}=10^{-8}$ (lower curve) and
      $3.33\times 10^{-7}$ (upper curve) and comparison with some theoretical
      predictions.}
  \label{fig:esp}
\end{figure}

These observations are studied in Fig.~\ref{fig:fc_gr} where the high
$\gamma \dot{F}$ limit is reached. At high damping the probability
distribution function $P(F)$ depends on $\dot{F}$ and $\gamma$ through
the $\gamma \dot{F}$ product. As seen in the figure, the
nonequilibrium high $\gamma \dot{F}$ regime appears in our simulations
for $\gamma \dot{F} > 10^{-4}$ ($\langle F \rangle > 0.13$ and $\Delta
U / k_BT < 2$). There we observe that an increasingly important
fraction of the realizations do not escape before getting the critical
force value ($V_0=0.155$) for which the potential profile does not
show a metastable well structure anymore. This situation is shown by
the filled circles curves in Fig.~\ref{fig:fc_gr} where the mean
escape force of the fraction of particles that have escaped from the
potential well for $F<V_0=0.155$ is plotted. An adequate theoretical
calculation based on the MFPT results allows for reproducing the
observed results even in this regime. To compare, we have also
computed the mean force for the particles to advance from the starting
point $a$ to a point $b=a+2\pi$. Result is show by open circles in the
figure and, as expected, the theoretical results based in MFPT theory
are also able to explain our results.

For the sake of completeness we present in Fig.~\ref{fig:esp} a plot
of the normalized mean escape force for a wide range of damping
values, $10^{-5}<\gamma<10^{3}$ and two different force rates. As
shown before, the escape rate has a maximum for $\gamma \sim 0.1$ and
thus the escape force presents a minimum for such damping value. For
larger values of $\gamma$ our Eq.~(\ref{rmhd}), correctly estimates
the escape force. For smaller values of $\gamma$, as shown
in~\cite{Mazo2010}, the DH equation for low damping
escape~\cite{DH1999} is valid. However, given the relative complexity
of both approaches, we want to compare the results to those obtained
with a simple interpolation formula, and the Garg analytical
predictions.

The simplest analytical expression which approach the escape rate for
all damping is given by the interpolation formula of
Eq.~(\ref{simple}). As seen in Fig.~\ref{fig:esp}, such expression
gives a very good approach to the computation of the mean escape force
and can be useful to analyze experiments at moderate damping values.
In addition, figure also shows the simple approximation published by
Garg (see also appendix). We plot first order approach for moderate
and high damping cases and second order for the low damping one. Such
results also give a good approach to our numerical results in a wide
damping area. It is important to remark now that as ramp is decreased,
the moderate damping area enlarges. Thus, the border between the
different damping regions is also given by the force ramp and a good
knowledge of both parameters is needed to correctly understand any
experimental result.

\section{Conclusions and discussion}

In this article we have presented a number of results concerning the
escape problem for moderate-to-high and high damping and its consequences in force
spectroscopy experiments. Such problems include for instance cell to
cell adhesion and molecular bond
experiments~\cite{Bell78,Dudko2003,Husson08,Freund09}, dissociation of
molecular complexes~\cite{Lin07}, mechanical unfolding of
proteins~\cite{Hyeon2003}, single-molecule pulling
experiments~\cite{Hummer03,Dudko06,Maitra10,Maitra11,Luccioli10,Suzuki10},
free-energy
reconstruction~\cite{Hummer01,Wales12,Harris07,Hummer10,Tapia-Rojo12},
and study of friction at the
nanoscale~\cite{Sang01,Evstigneev04,Evstigneev06,Barel10,Krylov08,Fajardo10,Vanossi12,Schirmeisen05,Sang08}.

Our work gives an accurate expression, Eq.~(\ref{rmhd}), for the
escape rate in the moderate-to-high and high damping domains and
allows for a quantitative evaluation of errors made in the use of
different approaches to the problem. This result combined with the DH
one for the moderate-to-low and low damping regime allows for an study
of the thermal activation problem over a parabolic barrier for any
barrier and damping and a comparison with other simpler theoretical
approaches. Then we have quantitatively study some issues concerning
the relation between mean first passage time and the very definition
of a escape rate. We have considered also the time to reach the
barrier problem and compute its ratio with the usual definition of
escape rate at different values of the damping. Activation theory at
high damping is a central concept in order to understand force
spectroscopy results. We have focused in comparing predictions from
the different available theories. Infinite barrier results are usually
assumed in literature without an estimation of the real barrier values
involved in each case. We have explored this issue and showed that
finite barrier effects do not importantly affect force spectroscopy
results though escape indeed may occur at very low barrier values
(Fig.~\ref{fig:mM}). This was also obtained in
Ref.~\onlinecite{Husson09}. We have also discussed the role of $\gamma
\dot{F}$ as the relevant parameter at high damping.

As discussed in section III and Fig.~\ref{fig:MFPT_x}, the very
concept of {\em escape rate} loss its significance if position initial
conditions play a role. This happens when the two relevant time
scales, equilibration at the metastable well and escape time are not
separate.  Separation of time scales is a central issue in the theory
that has been qualitatively discussed in most of the seminal
papers. Fig.~\ref{fig:MFPT_x} summarizes our quantitative study. Then,
it appears that the concept of an escape rate becomes questionable for
barrier heights below 3 (in units of $k_BT$). In relation to this,
Fig.~\ref{fig:hd} shows that theory correctly accounts for our
numerical expectations at lower barriers, but certainly it has been
obtained with well defined initial conditions. However,
Fig.~\ref{fig:mM} also shows a good agreement and there, see bottom
inset of the figure, escape happens at very low normalized mean
barriers. In this case force has been adiabatically increased and
results are not affected by the chosen initial conditions at $F=0$. To
finish, Fig.~\ref{fig:fc_gr} gives, in terms of the $\gamma \dot{F}$
product, the limit of validity of theory when applied to force
spectroscopy problems. This limit is certainly reached when the
separation of time scales concept fails.

Another important issue is the role played by the damping
parameter. This is a pretty difficult issue to evaluate in many real
systems, although the validity of the high damping approach is usually
assumed. We see that, on the one hand, high damping theories works
fine for dimensionless damping $\gamma/\omega>2$. On the other hand,
if the $\gamma \dot{F}$ product goes above $10^{-4}$, important
non-equilibrium effects appear and usual escape rate theories are not
valid any more. We have pointed out the existence of a high $\gamma
\dot{F}$ limit beyond which the escape problem is not well defined (see
Fig.~\ref{fig:fc_gr}).

Frequently, the main source of uncertainty in the problem is the
potential profile itself. Then, sometimes it is assumed $r \sim r_0
{\rm exp}(-\Delta U/k_B T)$ with $r_0$ some empirical prefactor. In
such cases, all interest is focused on the barrier value and the field
dependence of this barrier, and the {\em weak} $r_0$ dependence with
$F$ is neglected. However, in general, $r_0$ contains information
about the damping, the bias field and the barrier shape.

In adhesion and other similar problems, free energy is frequently
modeled by a tilted Morse-type potential. Then, the applied field
creates a barrier~\cite{Freund09}. However, next potential minimum is
sited at $+\infty$ and escape rate is vaguely defined in terms of
being far (but not too far) from the potential barrier. In addition,
in some force spectroscopy experiments, the measured maximal spring
force corresponds to a rupture event~\cite{Hummer03}. Then the
particle over barrier picture may fail and the problem can be studied
as time to reach a given force, position or a maximum of a potential
profile. Mean time to reach the maximum can be an useful concept for
some specific problems and sometimes, when a cusplike barrier is
assumed, it is the only meaningful one. However, since in many cases
the prefactor is not carefully considered, possible errors on choosing
the correct picture are unwittingly integrated in uncertainties on the
different parameters involved in the escape rate prefactor.  This is
an interesting problem that we are currently studying in depth.

In force friction experiments, mean friction force is computed from
the maximum force in each
stick-slip~\cite{Sang01,Evstigneev04,Schirmeisen05,Sang08} cycle. A
study using expressions for the escape rate beyond the usual Kramers
approach to the escape rate is lacking. 

Most of the previously discussed issues are also relevant when
analyzing single-molecule pulling
experiments~\cite{Tshiprut09,Luccioli10,Dudko06}. An interesting
question here is that of the reconstruction of free-energy
landscapes~\cite{Hummer01}. Moreover, in such, and other systems,
additional issues beyond the scope of this manuscript as diffusion in
a multidimensional landscape~\cite{Suzuki10,Best10} has a fundamental
importance.

\begin{acknowledgments}  We acknowledge F. Falo and R. Tapia-Rojo for the reading of the manuscript.
This work was supported by Spain MICINN, projects FIS2008-01240 and
FIS2011-25167, co-financed by FEDER funds. We also thanks funding by
Gobierno de Arag\'on.
\end{acknowledgments}

\begin{appendix}

\section{}

Here we will give some numerical expressions that have been used along the manuscript.

\subsection{Pollak and Talkner.}
Pollak and Talkner proposed~\cite{Pollak93} the following
expression for the escape rate:
\begin{equation}
r_{_{PT}}=k_{_{KMHD}} \times r_{_{TST}}^{\rm exact} \times
\left(1+\frac{f(\chi)}{x}\right)
\end{equation}
where $x=\Delta U / k_B T$ and $\chi$ the following function of
$\bar{\gamma} = \gamma / \omega_b$
\begin{equation}
\chi=\frac{(1+\bar{\gamma}^2/4)^{1/2}}{\bar{\gamma}/2}
\end{equation}
and


\begin{align}
f(\chi)=\frac{1}{36\chi} & \Big[ 2-3\chi-\frac{1}{2}(\chi+1)^3
  + \nonumber \\
  + & \frac{3}{2} \frac{(\chi-1)^2(\chi+1)(3\chi^2+12\chi+1)}{9\chi^2-1} \Big].
\end{align}


In the high damping limit $\chi \to 1$ and $f(\chi) \to -5/36$ then
\begin{equation}
r_{_{PT\_HD}}=\frac{\omega_b}{\gamma} \times r_{_{TST}}^{\rm exact} \times
\left(1-\frac{5}{36x}\right).
\end{equation}

In this theory it is necessary to compute the term
\begin{equation}
r_{_{TST}}^{\rm exact}= \left\{ \sqrt{2\pi \beta m} \
e^{\beta V(x_{max})} \ \int_{-\infty}^{x_{max}} dx \ e^{-\beta V(x)} \right\}^{-1}.
\label{tst_e}
\end{equation}

Note that in the high barrier limit $r_{_{TST}}^{\rm exact} \simeq
\frac{\omega_a}{2 \pi} \, {\rm e}^{- \Delta U/k_B T}:=r_{_{TST}}$, the usual
result given by the transition state theory and used in Eq.~(\ref{Kramers}).

\subsection{Drozdov}
The problem was also studied by Drozdov in a series of
papers~\cite{Drozdov99,DH1999,Drozdov00} where he proposed:
\begin{equation}
r_{_D}= k_{_{D}} \times r_{_{TST}}^{\rm exact}.
\end{equation}
Here
\begin{equation}
k_{_{D}}= \left[
1+\frac{\gamma^4}{\omega_e^4} \left(
1+\frac{4 \theta}{n \omega_e \gamma^2}
\right)^n
 \right]^{-1/4},
\end{equation}
being $n=8/7$ a good choice,
\begin{equation}
\omega_e = \sqrt{2\pi / \beta} \left[
\int_{-\infty}^{+\infty} dx \ e^{\beta V(x)}
 \right]^{-1},
\end{equation}
and
\begin{equation}
\theta = \frac{\omega_e^2 \ \beta^{3/2}}{\sqrt{2\pi}}
\int_{-\infty}^{+\infty} dx \ V'^2(x) \ e^{\beta V(x)}.
\end{equation}

\begin{figure}[]
    \centering{\includegraphics[width=0.47\textwidth]{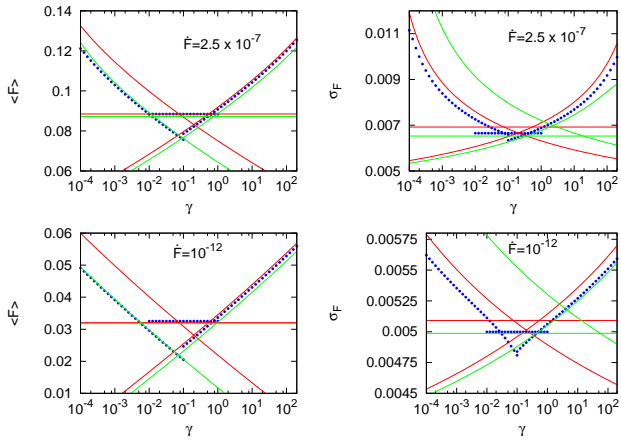}}
    \caption{(color online) Evaluation of Garg results for two
      different force ramps. Points stand for the exact numerical
      evaluation, solid lines for first order approximation, dashed
      lines for second order approximation results.}
  \label{fig:garg}
\end{figure}

\subsection{Garg}
To finish, we will present a short comparative study of
Garg~\cite{Garg} predictions for the mean escape force and its
variance. There, both quantities are given as a series expansion of a
certain parameter. Figure~\ref{fig:garg} shows a numerical study of
the results reported by Garg for two different force ramps. Barrier
height is given by $4\sqrt{2}V_0/3(1-F/V_0)^{3/2}$, oscillation
frequency by $2^{1/4}(V_0/m)^{1/2}(1-F/V_0)^{1/4}$, $T=0.01$ and
$\dot{F}=2.5 \times 10^{-7}$ and $10^{-12}$. Figure shows that for the
case of $\langle F \rangle$ first order approximation is very good for
the moderate and high damping cases (even better that considering the
second order approximation). On the contrary, in the low damping
result only the second order approximation correctly estimates the
exact escape rate. With respect to the variance, result are not that
good. For the case of moderate and low damping first order
approximation is not that bad and second order is better. However, for
low damping first order result is not very good but second order is
much worse.

\end{appendix}


\begin{thebibliography}{33}
\bibitem{comment2}{To normalize Eq.~(\ref{langevin}) we divide it by
  $V_0$ and time by $\omega^{-1}=\sqrt{m/V_0}$. Then, the
  dimensionless parameters are $\tilde{\gamma}=\gamma/\omega=1.50
  \gamma$, $\tilde{T}=k_B T / V_0=6.45 k_B T$ and
  $\widetilde{\dot{F}}=\dot{F} / (\omega V_0)=9.70 \dot{F}$.}
\end{thebibliography}
\end{document}